\documentclass[10pt,prl,twocolumn,showpacs,preprintnumbers,aps,superscriptaddress]{revtex4-1}
\usepackage{amsmath}
\usepackage{amssymb}
\usepackage{bm}
\usepackage{epsfig}
\usepackage{color}
\usepackage{comment}
\usepackage{float}
\usepackage{bm}
\usepackage{graphicx}

\usepackage{hyperref}
\usepackage[usenames,dvipsnames]{xcolor}

\usepackage{color}

\definecolor{darkgreen}{rgb}{0,0.6,0}

\definecolor{darkblue}{rgb}{0,0,0.6}

\definecolor{darkred}{rgb}{0.6,0,0}

\definecolor{darkpurple}{rgb}{0.5,0,0.5}

\hypersetup{
bookmarksopen=true,
pdftitle="",
pdfauthor="", 
pdftoolbar=false, 
pdfstartview={FitH},		
pdfmenubar=true,			
pdfhighlight=/O,			
colorlinks=true,			
urlcolor=darkblue,
citecolor=darkblue,		
linkcolor=darkblue}

\renewcommand{\phi}{\varphi}

\newcommand{\eps}{\epsilon}
\newcommand{\del}{\Delta}

\begin{document}

\title{Absence of Marginal Stability in a Structural Glass}

\author{Camille Scalliet}
\email{camille.scalliet@umontpellier.fr}
\author{Ludovic Berthier}

\affiliation{L2C, 
Universit\'e de Montpellier, CNRS, 34095 Montpellier, France}

\author{Francesco Zamponi}
\affiliation{Laboratoire de physique th\'eorique, D\'epartement de physique de
l'ENS, \'Ecole normale sup\'erieure, PSL
Research University, Sorbonne Universit\'es, UPMC Univ. Paris 06, CNRS,
75005 Paris, France}

\date{\today}

\begin{abstract}
Marginally stable solids have peculiar physical properties
that were first analyzed in the context of the 
jamming transition. 
We theoretically investigate the existence of marginal stability 
in a prototypical model for structural glass-formers, 
combining analytical calculations in infinite dimensions
to computer simulations in three dimensions.
While mean-field theory predicts the existence of
a Gardner phase transition towards a marginally 
stable glass phase at low temperatures,
simulations show no hint of diverging timescales or 
lengthscales, but reveal instead the presence of sparse localized defects. 
Our results suggest that the Gardner transition is deeply affected by finite
dimensional fluctuations, and raise issues about the relevance 
of marginal stability in structural glasses far away from jamming. 
\end{abstract}

\maketitle


Many types of fluids (molecular, colloidal, metallic)
transform into amorphous glasses~\cite{Ca09,BB11}. 
In the glass phase, they present thermodynamic~\cite{ZP71}, transport~\cite{Ph87},
vibrational~\cite{MS86} and mechanical~\cite{ML99} properties that are not observed in crystals.
These ``low-temperature anomalies'' are observed in a wide 
range of systems with very different particle types or interactions, 
and several theoretical approaches were developed to understand them~\cite{AHV72,Ph87,SDG98,LW01,LW06}, making specific assumptions about the nature of the excitations responsible for the anomalies. 

A different proposal recently emerged from the
convergence of two lines of research, based on the idea that collective excitations associated to marginal stability could be the key concept
underlying these properties. First, it was realized that systems close to a jamming transition 
are marginally stable, in the sense that 
the number of mechanical interactions in the system 
is precisely tuned~\cite{LNSW10}. It was later proposed that 
marginal stability persists away from 
jamming~\cite{WSNW05,XWLN07,DLDLW14}.
Second, an extension of the random first order transition 
theory~\cite{KW87,KT87,WL12} to 
amorphous hard spheres in large dimensions was obtained~\cite{CKPUZ16}. 
It predicts a Gardner
phase transition~\cite{GKS85,Ga85} between a normal glass phase and
a marginally stable one characterized by an excess of 
low-frequency modes~\cite{FPUZ15} 
and unusual rheological properties~\cite{BU16,FS17,JY17}. 
The marginal stability of the 
Gardner phase could provide a universal explanation 
for glass anomalies~\cite{MW15,CKPUZ16}.
{Marginal stability implies that the
system responds in a strong and system-spanning way to a weak, localized
perturbation~\cite{WSNW05}, implying the existence of 
delocalized soft modes~\cite{FPUZ15},
and diverging susceptibilities~\cite{HKLP11,BU16}. }

These recent results provide new opportunities to explain the properties of amorphous materials, motivating ongoing efforts to understand whether marginal stability holds generally in these materials.
Hard~\cite{BCJPSZ16,SD16,BR07} and soft~\cite{DLDLW14} spheres 
very close to the jamming transition have been analyzed,
showing that marginal stability and the Gardner transition may 
be relevant in colloidal and granular glasses. 
However, molecular and metallic glasses are usually modeled
by longer-ranged, continuous pair interactions
for which no jamming transition takes place~\cite{Ca09,BB11}.  In this context,
much less is known about the role of marginal stability~\cite{edan:2017}, 
and the existence of a Gardner phase has not been established.
Therefore, it is not known whether marginal stability 
can be used to understand the low-temperature anomalies in
generic structural glasses. 

To address this important question, we combined theoretical 
and numerical analysis of 
the low-temperature vibrational properties of a standard
model for atomic glasses. 
At the mean-field level, a marginally stable Gardner phase is predicted, 
which is then conceptually unrelated to jamming.
However, our numerical simulations of the same model in three dimensions
contrast with these predictions. We 
find no sign of a phase transition within the entire glass
phase. We detect instead sparse localized defects at 
low temperature, but they do not give rise to growing 
timescales and lengthscales that would accompany 
the emergence of marginal stability at a Gardner phase transition.


\textit{Mean-field theory} -- We consider a monodisperse system of $d$-dimensional particles interacting through a continuous pair potential $v(r)=\eps ( \sigma / r)^{4d}$. This is
the repulsive part of the Lennard-Jones potential, generalized to an arbitrary dimension $d$. The exponent for the inverse power law is larger than $d$ to ensure that the virial coefficients remain finite in any dimension.
We use $\sigma$ and $\eps$ as our unit 
length and energy, respectively. The state of the system is uniquely controlled 
by $\Gamma = \widehat \phi /T^{1/4}$, where $T$ is the temperature and 
$\widehat \phi = \rho V_ d 2^d  / d$ is the rescaled 
packing fraction ($\rho$ is the number density, and 
$V_d$ the volume of a $d$-dimensional sphere of 
diameter unity). We fix the 
packing fraction $\widehat \phi = 1$ and vary the temperature,
thus exploring the entire phase diagram.

\begin{figure}
\includegraphics[scale = 1]{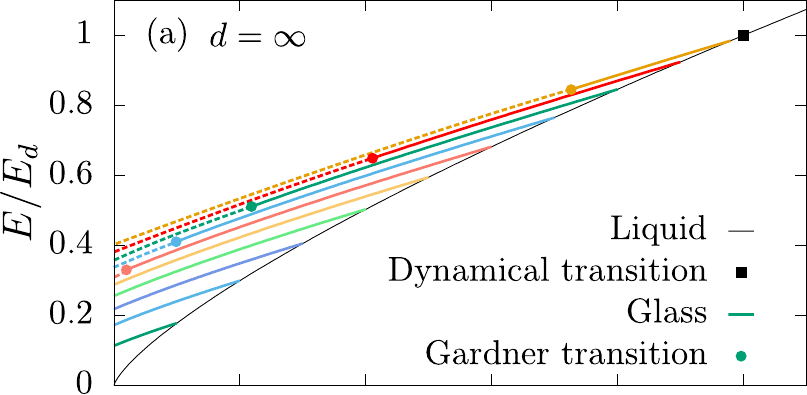}
\includegraphics[scale = 1]{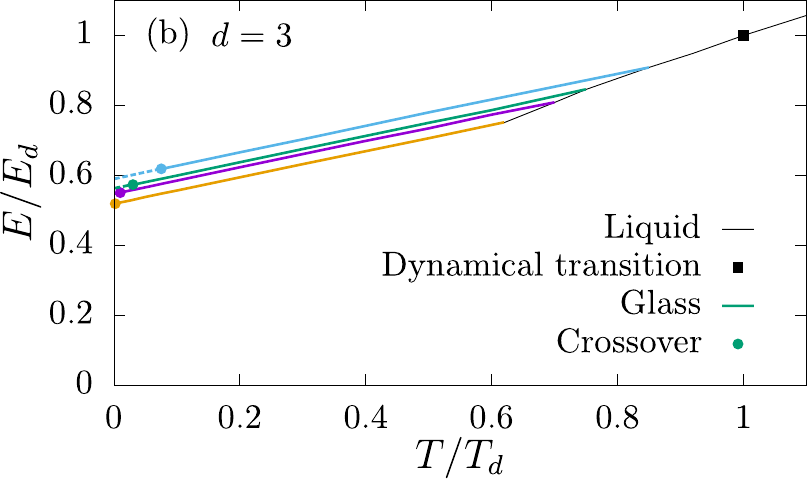}
\caption{(a) Mean-field phase diagram. Energy of the 
equilibrium liquid (black line), and dynamical transition temperature 
$T_d$ (black square). Various glasses prepared at 
$T_g < T_d$ are followed out of equilibrium (full lines).
When $T_g \gtrsim 0.5T_d$, these glasses undergo a Gardner transition
(bullets) into marginally stable glasses (dashed lines).
(b) Numerical phase diagram with equivalent representations of the liquid, 
dynamic transition and glass lines. Bullets locate the 
temperature crossover below which localized excitations appear, but do 
not correspond to a Gardner transition.
In both panels, axes are rescaled by the value at $T_d$: $E_d = E(T_d)$.}
\label{fig:MF}
\end{figure}

In the limit $d \rightarrow \infty$, the thermodynamic properties of the 
liquid and glass can be computed exactly~\cite{KW87,CKPUZ16}. 
In this limit, the system exhibits a sharp dynamical transition of the 
mode-coupling type~\cite{Go09,CKPUZ16} at a temperature $T_d$ at 
which the relaxation time of the liquid diverges. 
Below $T_d$, the system is trapped in one of the exponentially-many minima 
of the free-energy landscape. 
We compute the properties of a typical equilibrium liquid at a temperature 
$T_g \leq T_d$. As temperature decreases, the glass is confined near the state selected at $T_g$ in a ``restricted'' equilibrium, and thus follows an equation of state different from the liquid. We compute exactly the free energy of this
glass $f_g(T, T_g; \del, \del_r)$ at the 
replica symmetric level~\cite{RUYZ15} thanks to a state-following construction~\cite{FP95,RUYZ15,CKPUZ16}. 
It depends on two parameters: 
$\del$ is the long-time limit of the
mean-squared displacement within the followed 
glass state, and  
$\del_r$ is the relative mean-squared displacement 
between the original equilibrium configuration at $T_g$ and the one 
followed to $T$. The free energy $f_g$ is stationary 
with respect to $\del$ and $\del_r$. The average pressure and energy of glasses are obtained by taking derivatives of
 $f_g$ with respect to density and temperature, respectively. 

We solve the resulting set of coupled integro-differential 
equations given in Ref.~\cite{RUYZ15} to obtain 
the phase diagram in Fig.~\ref{fig:MF}a.
First, we compute the potential energy $E$ of the equilibrium liquid and the 
dynamical transition temperature, $T_d=0.002914$.
We then compute
the energy of glasses prepared at different $T_g \leq T_d$ as a function of temperature.
A Gardner transition is detected when the replica symmetric solution becomes 
unstable~\cite{RUYZ15},
signaling the transformation of the simple glass into a 
marginally stable one. The low-temperature Gardner phase is described by 
breaking the replica symmetry~\cite{Ga85,RU16}, 
and the transition belongs to the same universality class 
of the spin-glass transition in a magnetic field~\cite{MRT03,UB15}.
The presence of a Gardner transition is in general not a universal result~\cite{MRT03,Ri13}. In our model, 
over a large temperature window $0.5 \lesssim T_g/T_d <1$, a marginally 
stable Gardner phase exists,
while for $T_g \lesssim 0.5 T_d$ no Gardner transition is found. 
Because our model does not possess a jamming transition, our results show that
mean-field theory predicts that marginal stability 
is not restricted to the vicinity of jamming,
but should be broadly relevant for generic structural glasses with continuous 
interactions.


\textit{Numerical simulations} -- There is no clear consensus on the influence of finite dimensional fluctuations on the Gardner transition~\cite{LKMY13,Janus14,AB15,UB15,AKLMWY16,CY17}. Contradictory results were reported in numerical works. The existence of a transition was suggested in $d=4$~\cite{Janus14}, but opposite claims were also made~\cite{LKMY13,AKLMWY16}. A renormalization group approach~\cite{CY17} found a fixed point in all dimensions $d \geq 3$, while other works found different results~\cite{MB11,UB15}.
Thus, we must confront our theoretical predictions to a direct numerical investigation of the $3d$ version of the above model. Because the putative transition occurs deep inside the glass phase,
it is crucial to prepare well-thermalized glasses,
such that the structural relaxation time is larger than the 
duration of the simulation. This is now possible thanks to  
the development of an efficient swap Monte Carlo technique~\cite{GP01,Gutierrez:2015,NBC17}.

We simulate a continuously-polydisperse system composed of $N = 1500$ particles
at number density $\rho = 1$. We perform selected 
simulations with $N=12000$ to analyze finite size effects.
Particles interact via the repulsive pair 
potential $v(r_{ij}) = \eps (\sigma_{ij}/r_{ij})^{12}~+~F(r_{ij})$, 
where $F(r_{ij})$ guarantees the continuity of the potential up to the 
second derivative at the numerical cutoff distance $r_{cut} = 1.25~\sigma_{ij}$, 
beyond which $v=0$~\cite{NBC17}. The particle diameters are drawn from the normalized continuous distribution 
$P(\sigma_{m} \leq \sigma \leq \sigma_{M}) \sim1/\sigma^{3}$. 
The size ratio of $\sigma_{m} / \sigma_{M} = 0.45$ was optimized 
to provide an excellent glass-forming ability.
Similarly, we use a non-additive 
interaction rule for the cross diameters $\sigma_{ij} = \frac{\sigma_{i} + 
\sigma_{j}}{2}(1 - \eta|\sigma_{i} - \sigma_{j}|)$, with $\eta=0.2$.
Length, time and energy are 
respectively expressed in units of $\overline{\sigma} 
= \int  \sigma P(\sigma) d\sigma$, $\sqrt{\eps / m \overline{\sigma}^2}$ 
and $\epsilon$. The mode-coupling crossover temperature $T_d \approx 
0.1$ is determined by fitting the relaxation time $\tau$ measured with standard dynamics to $\tau \sim (T - T_d)^{- \gamma}$~\cite{NBC17}.
Using swap Monte Carlo, equilibrium can be ensured 
(using standard criteria \cite{NBC17}) down to $T \approx 0.6 T_d$. 
 
To numerically mimic the state following scheme, swap Monte Carlo is used to produce $N_{s}=50$ independent equilibrium configurations at each $T_g ~(0.062$, 0.07, 0.075, 0.082, 0.092). We then generate 
$N_{\rm th} = 10$ copies of each configuration, that differ only by the initial velocities of particles (two such copies are referred to as $A$ and $B$). 
Each of the $N_{s}\times N_{\rm th}$ samples 
is simulated at $T_g$ in the $NVE$ ensemble during a time $t_q$, depending on $T_g$ ($t_q = 1000$ for $T_g = 0.062, 0.07$;
$t_q = 100$ for $T_g = 0.075$; $t_q=0$ for $T_g=0.085$). 
The time $t_q$ is chosen such that particles in different copies have time to explore their cages without diffusing.
After $t_q$, the glass is instantaneously cooled to a temperature $T<T_g$ with a Berendsen thermostat (coupling parameter 
$\tau_T = 10$)~\cite{berendsen}. Waiting times $t_w$ are measured since the quench. We find that after $t_w \approx 100$, the temperature stabilizes to the desired value. For the highest $T_g = 0.092$ studied, diffusion is not totally suppressed at equilibrium. Glasses were first cooled down to $T = 0.07$ with a cooling rate $\gamma = 10^{-7}$ before making copies. We then used the same protocol as for $T_g = 0.07$ to obtain the data.

The Gardner transition is a second-order phase transition accompanied by diverging timescales and lengthscales characterizing vibrational dynamics. The transition signals profound changes in the structure of the landscape and the emergence of marginal stability. Mean-squared displacements (MSD) represent therefore the central observables for such investigation~\cite{CJPRSZ15,BCJPSZ16}:
\begin{equation}
\begin{split}
\Delta(t,t_w) &=\frac{1}{N'} \sum_{i=1}^{N'}  \langle | \boldsymbol{r}_i(t+t_w) - \boldsymbol{r}_i(t_w)|^2 \rangle~,  \\
\Delta_{AB}(t) &= \frac{1}{N'} \sum_{i=1}^{N'} \langle | \boldsymbol{r}_i^A(t) - \boldsymbol{r}_i^B(t)|^2 \rangle~,
\end{split}
\label{eq:msds}
\end{equation}
where the brackets indicate averages over 
thermal fluctuations and disorder. They 
respectively represent the standard MSD 
and the relative MSD between 
two copies of the same glass.
Since smaller particles may escape their cage more easily,
we concentrate on the $N' = N/2$ larger particles. 

\begin{figure}
\includegraphics[scale = 1.]{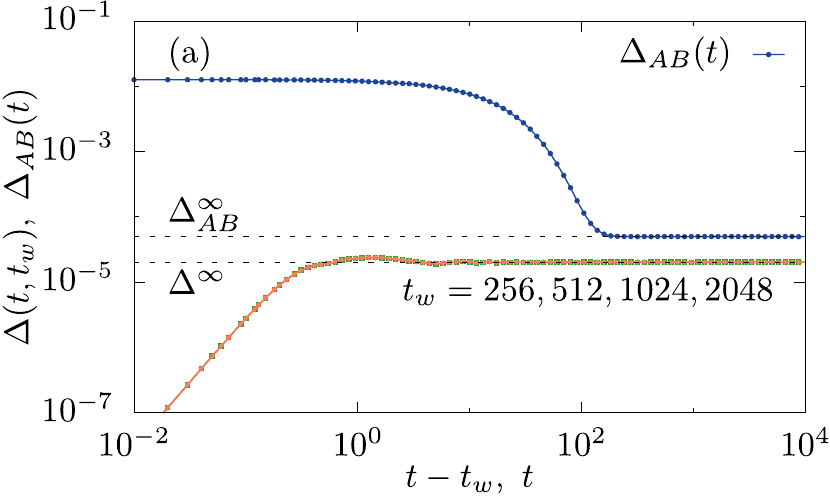}
\includegraphics[scale = 1.]{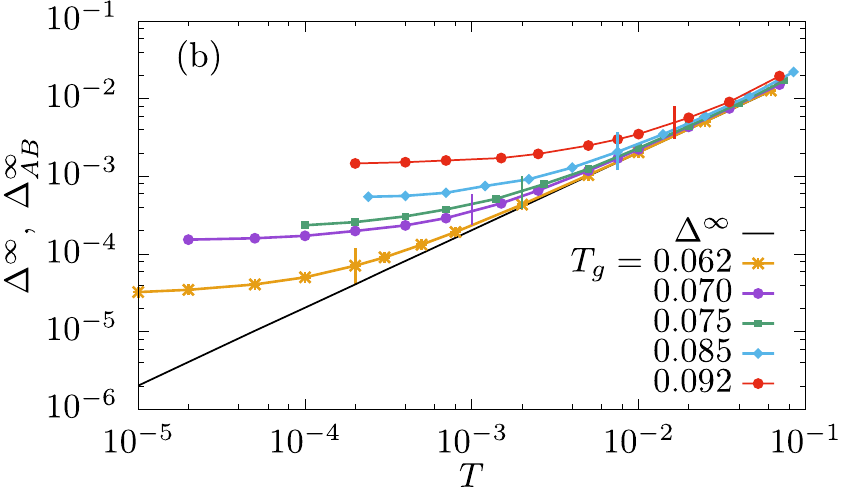}
\caption{(a) Mean-squared displacement $\Delta(t,t_w)$
and mean-squared distance $\del_{AB}(t)$ as a function of $t-t_w$ and 
$t$ for $T_g = 0.062$ and $T = 10^{-4}$ rapidly converge to their 
long-time limits (dashed lines).
(b) Long-time limits of $\del^{\infty}$ and 
$\del_{AB}^{\infty}$ as a function of temperature, for 
different $T_g$. Both quantities differ below $T^*(T_g)$,
indicated by vertical segments.}
\label{fig:deltasep}
\end{figure}

The typical behavior of the MSDs after a quench is shown in 
Fig.~\ref{fig:deltasep}a. Both quantities converge to their long-time limits,  $\del_{AB}^{\infty}$, $\del^{\infty}$ after a time 
of order 100 (set by the thermostat). No sign of slower relaxation or aging 
behavior is detected at any state point, which indicates that the time dependence of the observables is not pertinent. 
The absence of slow relaxation contrasts dramatically with 
hard sphere simulations~\cite{BCJPSZ16}, and directly reveals
the absence of marginal stability throughout the glass phase.

We gather the results for $\del_{AB}^{\infty}$ and 
$\del^{\infty}$ in Fig.~\ref{fig:deltasep}b. The standard MSD changes
linearly with $T$, as expected. The behavior of the relative distance 
is qualitatively the same for all $T_g$. The equality 
$\del_{AB}^{\infty} \approx \del^{\infty}$ holds at high enough $T$, 
meaning that the structure of the basin is 
relatively simple. There is a 
crossover temperature  $T^*(T_g)$ (vertical segments), below which 
$\del_{AB}^{\infty} > \del^{\infty}$. 
The distance between two copies is then much larger than the vibrations 
they can perform individually, suggesting that the copies get quenched
in distinct minima. This splitting of MSDs was observed in
hard spheres~\cite{BCJPSZ16,SD16} and identified as a Gardner transition. We report in Fig.~\ref{fig:MF}b the crossover temperatures 
and the glass energy. The similarity between the two phase diagrams in Fig. \ref{fig:MF} is obvious. 

\begin{figure}
\includegraphics[scale = 1.]{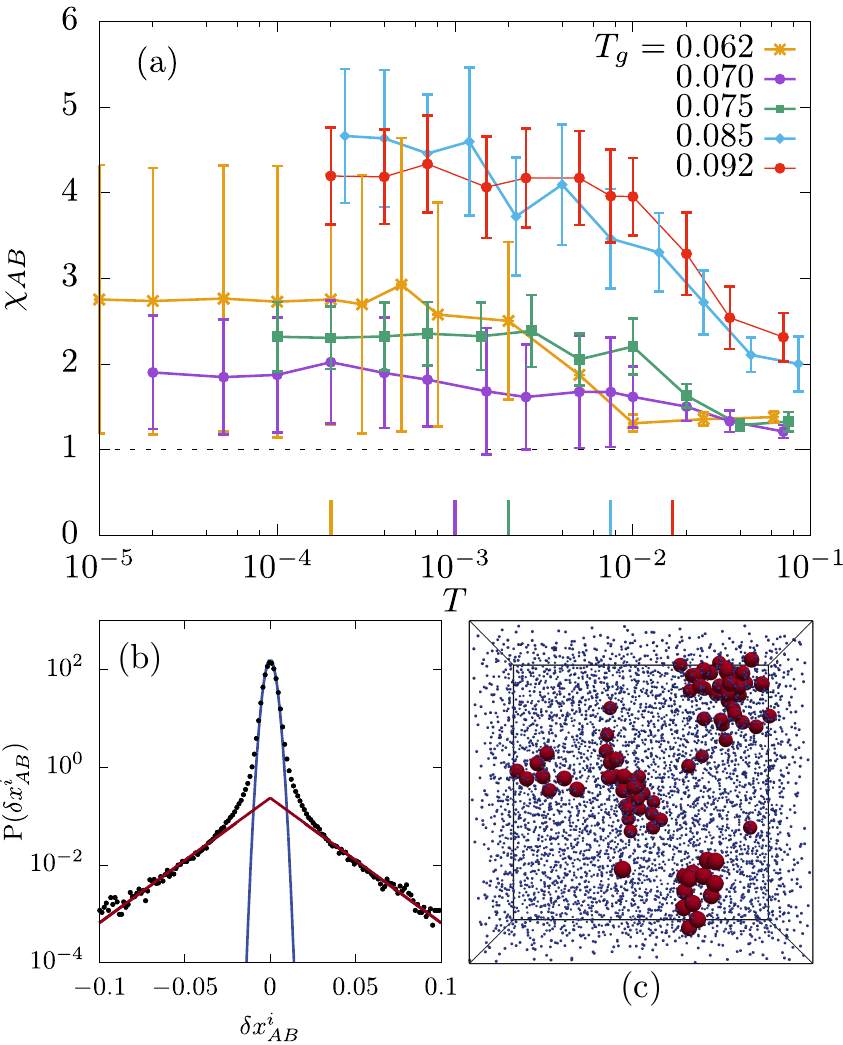}
\caption{(a): The susceptibility $\chi_{AB}$ for different $T_g$
raises mildly above the floor level (dashed line) across $T^*$ (vertical segments). The error bars are computed using the jackknife method.
(b): The van-Hove function of relative 
displacements has a narrow Gaussian core of width 
given by $\del^{\infty}$ 
(solid blue), and exponential tails (solid red)
at $T_g = 0.062$ and $T = 10^{-4}$.
(c) Corresponding snapshot with $N=12000$ showing the 
few particles having displacements outside the Gaussian range 
(red spheres) among a majority of 
particles undergoing small amplitude Gaussian vibrations (blue dots).}
\label{fig:pdfdispl}
\end{figure}

The absence of slow relaxation in Fig.~\ref{fig:deltasep}b reveals the lack of a growing timescale. To address lengthscales, we study the global fluctuations of the relative MSD. The variance of these fluctuations defines 
the susceptibility
$\chi_{AB} = N [ \langle \tilde{\Delta}_{AB}^2 \rangle - 
\langle \tilde{\Delta}_{AB} \rangle^2 ] / 
[ \langle \Delta_{AB}^{i~~~2} \rangle - \langle \Delta_{AB}^i\rangle^2 ]$, 
where $\tilde{\Delta}_{AB}$ 
is the plateau value of the relative MSD for a given pair $AB$,
and $\del^{i}_{AB}$ its single particle version.
The normalization in $\chi_{AB}$ ensures 
that $\chi_{AB}=1$ for spatially uncorrelated motion and that $\chi_{AB}$ is 
a direct measure of the correlation volume. If the crossover at $T^*$ corresponded to a Gardner transition, the susceptibility would diverge near $T^*$. The results in Fig.~\ref{fig:pdfdispl}a are very similar for 
all $T_g$ values: the susceptibility increases very weakly as temperature decreases. Within our error bars, 
there is actually very little global fluctuations above the floor level.
This directly demonstrates that spatial correlations between 
particle motion remain microscopic across the crossover $T^*$, 
which is thus not accompanied
by a growing correlation length. This is consistent
with the absence of slow dynamics in Fig.~\ref{fig:deltasep}a. 
A similar value of $\chi_{AB}$ was observed in larger systems of 
$N = 12 000$ particles at specific state points. We studied the spatial correlation function for the relative MSD~\cite{BCJPSZ16}, of which the volume integral is $\chi_{AB}$, and did not find hints of a growing length scale at any temperature.
We conclude that $T^*$ does not coincide with the emergence of a marginally 
stable phase. 

To understand the origin of the crossover observed in 
Fig.~\ref{fig:deltasep}b, we resolve the vibrational dynamics
at the particle scale. We measure the distribution 
of relative particle displacements, $P(\delta x_{AB}^i)$, 
where $\delta x_{AB}^i = x_A^i - x_B^i$ is the relative motion of
particle $i$ between copies $A$ and $B$ along the $x$-direction. We average 
over the three directions of space.
The van-Hove function is nearly Gaussian when $T^* < T < T_g$
with a width controlled by $\del^{\infty}$. 
Close to $T^*$ and below, the distribution remains Gaussian in its core, 
but exhibits tails that are well-fitted by an exponential, 
as shown in Fig.~\ref{fig:pdfdispl}b. 
We evaluate the statistical weight of the particles contributing to the 
tails by integrating the exponential fit. It varies between 1\% and 3\% 
for all state points and typically increases with $T_g$. 
This corresponds to a small subset of particles that 
get frozen in slightly distinct positions below $T^*$ in two copies.
The error bars for $\chi_{AB}$ are large because the number of particles in the tails is small, and fluctuates significantly from one pair $AB$ to another. We have checked that these mobile particles 
encompass all particles, not only small ones that are more mobile. 
To gather spatial information on these few mobile particles, 
we select particles with a relative displacement 
$\delta x_{AB}^i$ outside of the Gaussian core of the distribution, 
and visualize them in snapshots.
A typical snapshot obtained for $N = 12 000$ 
is shown in Fig.~\ref{fig:pdfdispl}c. We represent the 
vast majority of particles with Gaussian displacements as small points
and highlight particles contributing to the tails with larger
red spheres. 
Strikingly, these mobile particles are clustered into sparse localized defects. 
When $T_g$ increases, the number of mobile particles
as well as their characteristic displacement $\del^{i}_{AB}$ increase 
weakly. This directly accounts for the shift of $T^*$ with $T_g$.
These few localized clusters thus dominate the 
behavior of the relative displacement 
$\del_{AB}^{\infty}$ which is averaged over particles, and are responsible for its separation from $\del^{\infty}$
in Fig.~\ref{fig:deltasep}b. Our key conclusion is that 
the emergence of these localized clusters at $T^*$ does not
correspond to a Gardner phase transition, and glasses 
below $T^*$ are not marginally stable.\\


In our system, the marginal stability
described within mean-field approaches is strongly suppressed by
finite-dimensional fluctuations. Our 
results differ dramatically from previous work on 
hard sphere systems~\cite{BCJPSZ16,SD16}.  
This surprising lack of universality contrasts with the universality of 
glass formation~\cite{BB11}.
One possible explanation for this difference
is that structural glasses may generically become marginal only 
when pushed towards 
specific ``critical'' transitions, such as jamming~\cite{BU17}. 
The jamming 
transition appears robust down to $2d$, 
with the same critical properties as in $d=\infty$~\cite{GLN12}. 
This could explain why a Gardner transition is observed in 
finite-dimensional hard sphere glasses near jamming, 
whereas glass-formers with continuous interactions reach non-marginal inherent 
structures at zero temperature.
After this work was completed, two other works appeared reporting consistent findings~\cite{Moore:2017,Bea:2017}.

Our results raise two types of questions. First,
the presence of a marginally stable glass phase
seems highly dependent on 
the details of the particle interactions and on dimensionality.
To better understand the nature of the Gardner transition, one should 
investigate better the crossover between continuous and discontinuous 
interactions, using for instance well-chosen particle 
interactions~\cite{WCA71,BT11}.
One should also investigate the crossover towards non-mean-field 
behavior by using either dimensionality~\cite{CKPUZ16} or the interaction
range~\cite{MK11,CJPRSZ15} as tuning parameters.
Second, it would be interesting to connect the present 
results with other observations of localized defects~\cite{LW06}, 
such as soft localized modes controlling
the low-frequency part of the vibrational spectrum in 
amorphous solids~\cite{LDB16,MSI17},
{localized defects controlling relaxation 
in supercooled liquids~\cite{KHGGC11,CWKDBHR10,CSRMRDKL15,JG16},}
or the 
shear-transformation-zones~\cite{FL98,SWS07,HKLP10,PRB16} controlling 
the mechanical behavior of amorphous solids. 

\paragraph*{Acknowledgments --}
We thank G. Biroli, D. Coslovich, B. Seoane, P. Urbani for useful exchanges about this work, and A. Ninarello for providing initial configurations. The research leading to these results has received funding from the European Research Council under the European Unions Seventh Framework Programme (FP7/2007-2013)/ERC Grant Agreement No. 306845. This work was supported by a grant from the Simons Foundation (\#454933, Ludovic Berthier; \#454955, Francesco Zamponi).


\bibliographystyle{mioaps}

\bibliography{HS}

\begin{thebibliography}{65}
\expandafter\ifx\csname natexlab\endcsname\relax\def\natexlab#1{#1}\fi
\expandafter\ifx\csname bibnamefont\endcsname\relax
  \def\bibnamefont#1{#1}\fi
\expandafter\ifx\csname bibfnamefont\endcsname\relax
  \def\bibfnamefont#1{#1}\fi
\expandafter\ifx\csname citenamefont\endcsname\relax
  \def\citenamefont#1{#1}\fi
\expandafter\ifx\csname url\endcsname\relax
  \def\url#1{\texttt{#1}}\fi
\expandafter\ifx\csname urlprefix\endcsname\relax\def\urlprefix{URL }\fi
\providecommand{\bibinfo}[2]{#2}
\providecommand{\eprint}[2][]{\url{#2}}

\bibitem[{\citenamefont{Cavagna}(2009)}]{Ca09}
\bibinfo{author}{\bibfnamefont{A.}~\bibnamefont{Cavagna}},
  \bibinfo{journal}{Phys. Rep.} \textbf{\bibinfo{volume}{476}},
  \bibinfo{pages}{51} (\bibinfo{year}{2009}).

\bibitem[{\citenamefont{Berthier and Biroli}(2011)}]{BB11}
\bibinfo{author}{\bibfnamefont{L.}~\bibnamefont{Berthier}} \bibnamefont{and}
  \bibinfo{author}{\bibfnamefont{G.}~\bibnamefont{Biroli}},
  \bibinfo{journal}{Rev. Mod. Phys.} \textbf{\bibinfo{volume}{83}},
  \bibinfo{pages}{587} (\bibinfo{year}{2011}).

\bibitem[{\citenamefont{Zeller and Pohl}(1971)}]{ZP71}
\bibinfo{author}{\bibfnamefont{R.}~\bibnamefont{Zeller}} \bibnamefont{and}
  \bibinfo{author}{\bibfnamefont{R.}~\bibnamefont{Pohl}},
  \bibinfo{journal}{Phys. Rev. B} \textbf{\bibinfo{volume}{4}},
  \bibinfo{pages}{2029} (\bibinfo{year}{1971}).

\bibitem[{\citenamefont{Phillips}(1987)}]{Ph87}
\bibinfo{author}{\bibfnamefont{W.~A.} \bibnamefont{Phillips}},
  \bibinfo{journal}{Rep. Prog. Phys.} \textbf{\bibinfo{volume}{50}},
  \bibinfo{pages}{1657} (\bibinfo{year}{1987}).

\bibitem[{\citenamefont{Malinovsky and Sokolov}(1986)}]{MS86}
\bibinfo{author}{\bibfnamefont{V.~K.} \bibnamefont{Malinovsky}}
  \bibnamefont{and} \bibinfo{author}{\bibfnamefont{A.~P.}
  \bibnamefont{Sokolov}}, \bibinfo{journal}{Solid State Commun.}
  \textbf{\bibinfo{volume}{57}}, \bibinfo{pages}{757} (\bibinfo{year}{1986}).

\bibitem[{\citenamefont{Malandro and Lacks}(1999)}]{ML99}
\bibinfo{author}{\bibfnamefont{D.~L.} \bibnamefont{Malandro}} \bibnamefont{and}
  \bibinfo{author}{\bibfnamefont{D.~J.} \bibnamefont{Lacks}},
  \bibinfo{journal}{J. Chem. Phys.} \textbf{\bibinfo{volume}{110}},
  \bibinfo{pages}{4593} (\bibinfo{year}{1999}).

\bibitem[{\citenamefont{Anderson  \emph{et~al.}}(1972)\citenamefont{Anderson,
  Halperin, and Varma}}]{AHV72}
\bibinfo{author}{\bibfnamefont{P.~W.} \bibnamefont{Anderson}},
  \bibinfo{author}{\bibfnamefont{B.~I.} \bibnamefont{Halperin}},
  \bibnamefont{and} \bibinfo{author}{\bibfnamefont{C.~M.} \bibnamefont{Varma}},
  \bibinfo{journal}{Philos. Mag.} \textbf{\bibinfo{volume}{25}},
  \bibinfo{pages}{1} (\bibinfo{year}{1972}).

\bibitem[{\citenamefont{Schirmacher
  \emph{et~al.}}(1998)\citenamefont{Schirmacher, Diezemann, and
  Ganter}}]{SDG98}
\bibinfo{author}{\bibfnamefont{W.}~\bibnamefont{Schirmacher}},
  \bibinfo{author}{\bibfnamefont{G.}~\bibnamefont{Diezemann}},
  \bibnamefont{and} \bibinfo{author}{\bibfnamefont{C.}~\bibnamefont{Ganter}},
  \bibinfo{journal}{Phys. Rev. Lett.} \textbf{\bibinfo{volume}{81}},
  \bibinfo{pages}{136} (\bibinfo{year}{1998}).

\bibitem[{\citenamefont{Lubchenko and Wolynes}(2001)}]{LW01}
\bibinfo{author}{\bibfnamefont{V.}~\bibnamefont{Lubchenko}} \bibnamefont{and}
  \bibinfo{author}{\bibfnamefont{P.~G.} \bibnamefont{Wolynes}},
  \bibinfo{journal}{Phys. Rev. Lett.} \textbf{\bibinfo{volume}{87}},
  \bibinfo{pages}{195901} (\bibinfo{year}{2001}).

\bibitem[{\citenamefont{Lubchenko and Wolynes}(2007)}]{LW06}
\bibinfo{author}{\bibfnamefont{V.}~\bibnamefont{Lubchenko}} \bibnamefont{and}
  \bibinfo{author}{\bibfnamefont{P.~G.} \bibnamefont{Wolynes}},
  \bibinfo{journal}{Annu. Rev. Phys. Chem.} \textbf{\bibinfo{volume}{58}},
  \bibinfo{pages}{235} (\bibinfo{year}{2007}), \bibinfo{note}{pMID: 17067282}.

\bibitem[{\citenamefont{Liu  \emph{et~al.}}(2011)\citenamefont{Liu, Nagel,
  Van~Saarloos, and Wyart}}]{LNSW10}
\bibinfo{author}{\bibfnamefont{A.}~\bibnamefont{Liu}},
  \bibinfo{author}{\bibfnamefont{S.}~\bibnamefont{Nagel}},
  \bibinfo{author}{\bibfnamefont{W.}~\bibnamefont{Van~Saarloos}},
  \bibnamefont{and} \bibinfo{author}{\bibfnamefont{M.}~\bibnamefont{Wyart}}, in
  \emph{\bibinfo{booktitle}{Dynamical Heterogeneities and Glasses}}, edited by
  \bibinfo{editor}{\bibfnamefont{L.}~\bibnamefont{Berthier}},
  \bibinfo{editor}{\bibfnamefont{G.}~\bibnamefont{Biroli}},
  \bibinfo{editor}{\bibfnamefont{J.-P.} \bibnamefont{Bouchaud}},
  \bibinfo{editor}{\bibfnamefont{L.}~\bibnamefont{Cipelletti}},
  \bibnamefont{and} \bibinfo{editor}{\bibfnamefont{W.}~\bibnamefont{van
  Saarloos}} (\bibinfo{publisher}{Oxford University Press, Oxford},
  \bibinfo{year}{2011}), \eprint{{\tt arXiv:1006.2365}}.

\bibitem[{\citenamefont{Wyart  \emph{et~al.}}(2005)\citenamefont{Wyart,
  Silbert, Nagel, and Witten}}]{WSNW05}
\bibinfo{author}{\bibfnamefont{M.}~\bibnamefont{Wyart}},
  \bibinfo{author}{\bibfnamefont{L.}~\bibnamefont{Silbert}},
  \bibinfo{author}{\bibfnamefont{S.}~\bibnamefont{Nagel}}, \bibnamefont{and}
  \bibinfo{author}{\bibfnamefont{T.}~\bibnamefont{Witten}},
  \bibinfo{journal}{Phys. Rev. E} \textbf{\bibinfo{volume}{72}},
  \bibinfo{pages}{051306} (\bibinfo{year}{2005}).

\bibitem[{\citenamefont{Xu  \emph{et~al.}}(2007)\citenamefont{Xu, Wyart, Liu,
  and Nagel}}]{XWLN07}
\bibinfo{author}{\bibfnamefont{N.}~\bibnamefont{Xu}},
  \bibinfo{author}{\bibfnamefont{M.}~\bibnamefont{Wyart}},
  \bibinfo{author}{\bibfnamefont{A.~J.} \bibnamefont{Liu}}, \bibnamefont{and}
  \bibinfo{author}{\bibfnamefont{S.~R.} \bibnamefont{Nagel}},
  \bibinfo{journal}{Phys. Rev. Lett.} \textbf{\bibinfo{volume}{98}},
  \bibinfo{pages}{175502} (\bibinfo{year}{2007}).

\bibitem[{\citenamefont{DeGiuli  \emph{et~al.}}(2014)\citenamefont{DeGiuli,
  Laversanne-Finot, D{\"u}ring, Lerner, and Wyart}}]{DLDLW14}
\bibinfo{author}{\bibfnamefont{E.}~\bibnamefont{DeGiuli}},
  \bibinfo{author}{\bibfnamefont{A.}~\bibnamefont{Laversanne-Finot}},
  \bibinfo{author}{\bibfnamefont{G.}~\bibnamefont{D{\"u}ring}},
  \bibinfo{author}{\bibfnamefont{E.}~\bibnamefont{Lerner}}, \bibnamefont{and}
  \bibinfo{author}{\bibfnamefont{M.}~\bibnamefont{Wyart}},
  \bibinfo{journal}{Soft Matter} \textbf{\bibinfo{volume}{10}},
  \bibinfo{pages}{5628} (\bibinfo{year}{2014}).

\bibitem[{\citenamefont{Kirkpatrick and Wolynes}(1987)}]{KW87}
\bibinfo{author}{\bibfnamefont{T.~R.} \bibnamefont{Kirkpatrick}}
  \bibnamefont{and} \bibinfo{author}{\bibfnamefont{P.~G.}
  \bibnamefont{Wolynes}}, \bibinfo{journal}{Phys. Rev. A}
  \textbf{\bibinfo{volume}{35}}, \bibinfo{pages}{3072} (\bibinfo{year}{1987}).

\bibitem[{\citenamefont{Kirkpatrick and Thirumalai}(1987)}]{KT87}
\bibinfo{author}{\bibfnamefont{T.~R.} \bibnamefont{Kirkpatrick}}
  \bibnamefont{and}
  \bibinfo{author}{\bibfnamefont{D.}~\bibnamefont{Thirumalai}},
  \bibinfo{journal}{Phys. Rev. Lett.} \textbf{\bibinfo{volume}{58}},
  \bibinfo{pages}{2091} (\bibinfo{year}{1987}).

\bibitem[{\citenamefont{Lubchenko and Wolynes}(2012)}]{WL12}
\bibinfo{editor}{\bibfnamefont{V.}~\bibnamefont{Lubchenko}} \bibnamefont{and}
  \bibinfo{editor}{\bibfnamefont{P.~G.} \bibnamefont{Wolynes}}, eds.,
  \emph{\bibinfo{title}{Structural Glasses and Supercooled Liquids: Theory,
  Experiment, and Applications}} (\bibinfo{publisher}{Wiley},
  \bibinfo{year}{2012}).

\bibitem[{\citenamefont{Charbonneau
  \emph{et~al.}}(2017)\citenamefont{Charbonneau, Kurchan, Parisi, Urbani, and
  Zamponi}}]{CKPUZ16}
\bibinfo{author}{\bibfnamefont{P.}~\bibnamefont{Charbonneau}},
  \bibinfo{author}{\bibfnamefont{J.}~\bibnamefont{Kurchan}},
  \bibinfo{author}{\bibfnamefont{G.}~\bibnamefont{Parisi}},
  \bibinfo{author}{\bibfnamefont{P.}~\bibnamefont{Urbani}}, \bibnamefont{and}
  \bibinfo{author}{\bibfnamefont{F.}~\bibnamefont{Zamponi}},
  \bibinfo{journal}{Annu. Rev. Condens. Matter Phys.}
  \textbf{\bibinfo{volume}{8}}, \bibinfo{pages}{265} (\bibinfo{year}{2017}).

\bibitem[{\citenamefont{Gross  \emph{et~al.}}(1985)\citenamefont{Gross, Kanter,
  and Sompolinsky}}]{GKS85}
\bibinfo{author}{\bibfnamefont{D.~J.} \bibnamefont{Gross}},
  \bibinfo{author}{\bibfnamefont{I.}~\bibnamefont{Kanter}}, \bibnamefont{and}
  \bibinfo{author}{\bibfnamefont{H.}~\bibnamefont{Sompolinsky}},
  \bibinfo{journal}{Phys. Rev. Lett.} \textbf{\bibinfo{volume}{55}},
  \bibinfo{pages}{304} (\bibinfo{year}{1985}).

\bibitem[{\citenamefont{Gardner}(1985)}]{Ga85}
\bibinfo{author}{\bibfnamefont{E.}~\bibnamefont{Gardner}},
  \bibinfo{journal}{Nucl. Phys. B} \textbf{\bibinfo{volume}{257}},
  \bibinfo{pages}{747} (\bibinfo{year}{1985}).

\bibitem[{\citenamefont{Franz  \emph{et~al.}}(2015)\citenamefont{Franz, Parisi,
  Urbani, and Zamponi}}]{FPUZ15}
\bibinfo{author}{\bibfnamefont{S.}~\bibnamefont{Franz}},
  \bibinfo{author}{\bibfnamefont{G.}~\bibnamefont{Parisi}},
  \bibinfo{author}{\bibfnamefont{P.}~\bibnamefont{Urbani}}, \bibnamefont{and}
  \bibinfo{author}{\bibfnamefont{F.}~\bibnamefont{Zamponi}},
  \bibinfo{journal}{Proc. Natl. Acad. Sci. U.S.A.}
  \textbf{\bibinfo{volume}{112}}, \bibinfo{pages}{14539}
  (\bibinfo{year}{2015}).

\bibitem[{\citenamefont{Biroli and Urbani}(2016)}]{BU16}
\bibinfo{author}{\bibfnamefont{G.}~\bibnamefont{Biroli}} \bibnamefont{and}
  \bibinfo{author}{\bibfnamefont{P.}~\bibnamefont{Urbani}},
  \bibinfo{journal}{Nature Physics} \textbf{\bibinfo{volume}{12}},
  \bibinfo{pages}{1130} (\bibinfo{year}{2016}).

\bibitem[{\citenamefont{Franz and Spigler}(2017)}]{FS17}
\bibinfo{author}{\bibfnamefont{S.}~\bibnamefont{Franz}} \bibnamefont{and}
  \bibinfo{author}{\bibfnamefont{S.}~\bibnamefont{Spigler}},
  \bibinfo{journal}{Phys. Rev. E} \textbf{\bibinfo{volume}{95}},
  \bibinfo{pages}{022139} (\bibinfo{year}{2017}).

\bibitem[{\citenamefont{Jin and Yoshino}(2017)}]{JY17}
\bibinfo{author}{\bibfnamefont{Y.}~\bibnamefont{Jin}} \bibnamefont{and}
  \bibinfo{author}{\bibfnamefont{H.}~\bibnamefont{Yoshino}},
  \bibinfo{journal}{Nature Communications} \textbf{\bibinfo{volume}{8}},
  \bibinfo{pages}{14935} (\bibinfo{year}{2017}).

\bibitem[{\citenamefont{M{\"u}ller and Wyart}(2015)}]{MW15}
\bibinfo{author}{\bibfnamefont{M.}~\bibnamefont{M{\"u}ller}} \bibnamefont{and}
  \bibinfo{author}{\bibfnamefont{M.}~\bibnamefont{Wyart}},
  \bibinfo{journal}{Annu. Rev. Condens. Matter Phys.}
  \textbf{\bibinfo{volume}{6}}, \bibinfo{pages}{177} (\bibinfo{year}{2015}).

\bibitem[{\citenamefont{Hentschel  \emph{et~al.}}(2011)\citenamefont{Hentschel,
  Karmakar, Lerner, and Procaccia}}]{HKLP11}
\bibinfo{author}{\bibfnamefont{H.}~\bibnamefont{Hentschel}},
  \bibinfo{author}{\bibfnamefont{S.}~\bibnamefont{Karmakar}},
  \bibinfo{author}{\bibfnamefont{E.}~\bibnamefont{Lerner}}, \bibnamefont{and}
  \bibinfo{author}{\bibfnamefont{I.}~\bibnamefont{Procaccia}},
  \bibinfo{journal}{Phys. Rev. E} \textbf{\bibinfo{volume}{83}},
  \bibinfo{pages}{061101} (\bibinfo{year}{2011}).

\bibitem[{\citenamefont{Berthier  \emph{et~al.}}(2016)\citenamefont{Berthier,
  Charbonneau, Jin, Parisi, Seoane, and Zamponi}}]{BCJPSZ16}
\bibinfo{author}{\bibfnamefont{L.}~\bibnamefont{Berthier}},
  \bibinfo{author}{\bibfnamefont{P.}~\bibnamefont{Charbonneau}},
  \bibinfo{author}{\bibfnamefont{Y.}~\bibnamefont{Jin}},
  \bibinfo{author}{\bibfnamefont{G.}~\bibnamefont{Parisi}},
  \bibinfo{author}{\bibfnamefont{B.}~\bibnamefont{Seoane}}, \bibnamefont{and}
  \bibinfo{author}{\bibfnamefont{F.}~\bibnamefont{Zamponi}},
  \bibinfo{journal}{Proc. Natl. Acad. Sci. U.S.A.}
  \textbf{\bibinfo{volume}{113}}, \bibinfo{pages}{8397} (\bibinfo{year}{2016}).

\bibitem[{\citenamefont{Seguin and Dauchot}(2016)}]{SD16}
\bibinfo{author}{\bibfnamefont{A.}~\bibnamefont{Seguin}} \bibnamefont{and}
  \bibinfo{author}{\bibfnamefont{O.}~\bibnamefont{Dauchot}},
  \bibinfo{journal}{Phys. Rev. Lett.} \textbf{\bibinfo{volume}{117}},
  \bibinfo{pages}{228001} (\bibinfo{year}{2016}).

\bibitem[{\citenamefont{Brito and Wyart}(2007)}]{BR07}
\bibinfo{author}{\bibfnamefont{C.}~\bibnamefont{Brito}} \bibnamefont{and}
  \bibinfo{author}{\bibfnamefont{M.}~\bibnamefont{Wyart}}, \bibinfo{journal}{J.
  Stat. Mech.} \textbf{\bibinfo{volume}{2007}}, \bibinfo{pages}{L08003}
  (\bibinfo{year}{2007}).

\bibitem[{\citenamefont{Kooij and Lerner}(2017)}]{edan:2017}
\bibinfo{author}{\bibfnamefont{S.}~\bibnamefont{Kooij}} \bibnamefont{and}
  \bibinfo{author}{\bibfnamefont{E.}~\bibnamefont{Lerner}},
  \bibinfo{journal}{Phys. Rev. E} \textbf{\bibinfo{volume}{95}},
  \bibinfo{pages}{062141} (\bibinfo{year}{2017}).

\bibitem[{\citenamefont{G{\"o}tze}(2009)}]{Go09}
\bibinfo{author}{\bibfnamefont{W.}~\bibnamefont{G{\"o}tze}},
  \emph{\bibinfo{title}{Complex dynamics of glass-forming liquids: A
  mode-coupling theory}}, vol. \bibinfo{volume}{143} (\bibinfo{publisher}{OUP,
  USA}, \bibinfo{year}{2009}).

\bibitem[{\citenamefont{Rainone  \emph{et~al.}}(2015)\citenamefont{Rainone,
  Urbani, Yoshino, and Zamponi}}]{RUYZ15}
\bibinfo{author}{\bibfnamefont{C.}~\bibnamefont{Rainone}},
  \bibinfo{author}{\bibfnamefont{P.}~\bibnamefont{Urbani}},
  \bibinfo{author}{\bibfnamefont{H.}~\bibnamefont{Yoshino}}, \bibnamefont{and}
  \bibinfo{author}{\bibfnamefont{F.}~\bibnamefont{Zamponi}},
  \bibinfo{journal}{Phys. Rev. Lett.} \textbf{\bibinfo{volume}{114}},
  \bibinfo{pages}{015701} (\bibinfo{year}{2015}).

\bibitem[{\citenamefont{{Silvio Franz} and {Giorgio Parisi}}(1995)}]{FP95}
\bibinfo{author}{\bibnamefont{{Silvio Franz}}} \bibnamefont{and}
  \bibinfo{author}{\bibnamefont{{Giorgio Parisi}}}, \bibinfo{journal}{J. Phys.
  I France} \textbf{\bibinfo{volume}{5}}, \bibinfo{pages}{1401}
  (\bibinfo{year}{1995}).

\bibitem[{\citenamefont{Rainone and Urbani}(2016)}]{RU16}
\bibinfo{author}{\bibfnamefont{C.}~\bibnamefont{Rainone}} \bibnamefont{and}
  \bibinfo{author}{\bibfnamefont{P.}~\bibnamefont{Urbani}},
  \bibinfo{journal}{J. Stat Mech.} \textbf{\bibinfo{volume}{2016}},
  \bibinfo{pages}{P053302} (\bibinfo{year}{2016}).

\bibitem[{\citenamefont{Montanari and Ricci-Tersenghi}(2003)}]{MRT03}
\bibinfo{author}{\bibfnamefont{A.}~\bibnamefont{Montanari}} \bibnamefont{and}
  \bibinfo{author}{\bibfnamefont{F.}~\bibnamefont{Ricci-Tersenghi}},
  \bibinfo{journal}{Eur. Phys. J. B} \textbf{\bibinfo{volume}{33}},
  \bibinfo{pages}{339} (\bibinfo{year}{2003}).

\bibitem[{\citenamefont{Urbani and Biroli}(2015)}]{UB15}
\bibinfo{author}{\bibfnamefont{P.}~\bibnamefont{Urbani}} \bibnamefont{and}
  \bibinfo{author}{\bibfnamefont{G.}~\bibnamefont{Biroli}},
  \bibinfo{journal}{Phys. Rev. B} \textbf{\bibinfo{volume}{91}},
  \bibinfo{pages}{100202} (\bibinfo{year}{2015}).

\bibitem[{\citenamefont{Rizzo}(2013)}]{Ri13}
\bibinfo{author}{\bibfnamefont{T.}~\bibnamefont{Rizzo}},
  \bibinfo{journal}{Phys. Rev. E} \textbf{\bibinfo{volume}{88}},
  \bibinfo{pages}{032135} (\bibinfo{year}{2013}).

\bibitem[{\citenamefont{Larson  \emph{et~al.}}(2013)\citenamefont{Larson,
  Katzgraber, Moore, and Young}}]{LKMY13}
\bibinfo{author}{\bibfnamefont{D.}~\bibnamefont{Larson}},
  \bibinfo{author}{\bibfnamefont{H.~G.} \bibnamefont{Katzgraber}},
  \bibinfo{author}{\bibfnamefont{M.~A.} \bibnamefont{Moore}}, \bibnamefont{and}
  \bibinfo{author}{\bibfnamefont{A.~P.} \bibnamefont{Young}},
  \bibinfo{journal}{Phys. Rev. B} \textbf{\bibinfo{volume}{87}},
  \bibinfo{pages}{024414} (\bibinfo{year}{2013}).

\bibitem[{\citenamefont{Baity-Jesi  \emph{et~al.}}(2014)}]{Janus14}
\bibinfo{author}{\bibfnamefont{M.}~\bibnamefont{Baity-Jesi}} \bibnamefont{
  \emph{et~al.}}, \bibinfo{journal}{J. Stat. Mech.}
  \textbf{\bibinfo{volume}{2014}}, \bibinfo{pages}{P05014}
  (\bibinfo{year}{2014}).

\bibitem[{\citenamefont{Angelini and Biroli}(2015)}]{AB15}
\bibinfo{author}{\bibfnamefont{M.~C.} \bibnamefont{Angelini}} \bibnamefont{and}
  \bibinfo{author}{\bibfnamefont{G.}~\bibnamefont{Biroli}},
  \bibinfo{journal}{Phys. Rev. Lett.} \textbf{\bibinfo{volume}{114}},
  \bibinfo{pages}{095701} (\bibinfo{year}{2015}).

\bibitem[{\citenamefont{Aspelmeier
  \emph{et~al.}}(2016)\citenamefont{Aspelmeier, Katzgraber, Larson, Moore,
  Wittmann, and Yeo}}]{AKLMWY16}
\bibinfo{author}{\bibfnamefont{T.}~\bibnamefont{Aspelmeier}},
  \bibinfo{author}{\bibfnamefont{H.~G.} \bibnamefont{Katzgraber}},
  \bibinfo{author}{\bibfnamefont{D.}~\bibnamefont{Larson}},
  \bibinfo{author}{\bibfnamefont{M.~A.} \bibnamefont{Moore}},
  \bibinfo{author}{\bibfnamefont{M.}~\bibnamefont{Wittmann}}, \bibnamefont{and}
  \bibinfo{author}{\bibfnamefont{J.}~\bibnamefont{Yeo}},
  \bibinfo{journal}{Phys. Rev. E} \textbf{\bibinfo{volume}{93}},
  \bibinfo{pages}{032123} (\bibinfo{year}{2016}).

\bibitem[{\citenamefont{Charbonneau and Yaida}(2017)}]{CY17}
\bibinfo{author}{\bibfnamefont{P.}~\bibnamefont{Charbonneau}} \bibnamefont{and}
  \bibinfo{author}{\bibfnamefont{S.}~\bibnamefont{Yaida}},
  \bibinfo{journal}{Phys. Rev. Lett.} \textbf{\bibinfo{volume}{118}},
  \bibinfo{pages}{215701} (\bibinfo{year}{2017}).

\bibitem[{\citenamefont{Moore and Bray}(2011)}]{MB11}
\bibinfo{author}{\bibfnamefont{M.~A.} \bibnamefont{Moore}} \bibnamefont{and}
  \bibinfo{author}{\bibfnamefont{A.~J.} \bibnamefont{Bray}},
  \bibinfo{journal}{Phys. Rev. B} \textbf{\bibinfo{volume}{83}},
  \bibinfo{pages}{224408} (\bibinfo{year}{2011}).

\bibitem[{\citenamefont{Grigera and Parisi}(2001)}]{GP01}
\bibinfo{author}{\bibfnamefont{T.}~\bibnamefont{Grigera}} \bibnamefont{and}
  \bibinfo{author}{\bibfnamefont{G.}~\bibnamefont{Parisi}},
  \bibinfo{journal}{Phys. Rev. E} \textbf{\bibinfo{volume}{63}},
  \bibinfo{pages}{45102} (\bibinfo{year}{2001}).

\bibitem[{\citenamefont{Guti\'errez
  \emph{et~al.}}(2015)\citenamefont{Guti\'errez, Karmakar, Pollack, and
  Procaccia}}]{Gutierrez:2015}
\bibinfo{author}{\bibfnamefont{R.}~\bibnamefont{Guti\'errez}},
  \bibinfo{author}{\bibfnamefont{S.}~\bibnamefont{Karmakar}},
  \bibinfo{author}{\bibfnamefont{Y.~G.} \bibnamefont{Pollack}},
  \bibnamefont{and}
  \bibinfo{author}{\bibfnamefont{I.}~\bibnamefont{Procaccia}},
  \bibinfo{journal}{Europhys. Lett.} \textbf{\bibinfo{volume}{111}},
  \bibinfo{pages}{56009} (\bibinfo{year}{2015}).

\bibitem[{\citenamefont{Ninarello  \emph{et~al.}}(2017)\citenamefont{Ninarello,
  Berthier, and Coslovich}}]{NBC17}
\bibinfo{author}{\bibfnamefont{A.}~\bibnamefont{Ninarello}},
  \bibinfo{author}{\bibfnamefont{L.}~\bibnamefont{Berthier}}, \bibnamefont{and}
  \bibinfo{author}{\bibfnamefont{D.}~\bibnamefont{Coslovich}},
  \bibinfo{journal}{Phys. Rev. X} \textbf{\bibinfo{volume}{7}},
  \bibinfo{pages}{021039} (\bibinfo{year}{2017}).

\bibitem[{\citenamefont{Berendsen  \emph{et~al.}}(1984)\citenamefont{Berendsen,
  Postma, van Gunsteren, DiNola, and Haak}}]{berendsen}
\bibinfo{author}{\bibfnamefont{H.~J.~C.} \bibnamefont{Berendsen}},
  \bibinfo{author}{\bibfnamefont{J.~P.~M.} \bibnamefont{Postma}},
  \bibinfo{author}{\bibfnamefont{W.~F.} \bibnamefont{van Gunsteren}},
  \bibinfo{author}{\bibfnamefont{A.}~\bibnamefont{DiNola}}, \bibnamefont{and}
  \bibinfo{author}{\bibfnamefont{J.~R.} \bibnamefont{Haak}},
  \bibinfo{journal}{J. Chem. Phys.} \textbf{\bibinfo{volume}{81}},
  \bibinfo{pages}{3684} (\bibinfo{year}{1984}).

\bibitem[{\citenamefont{Charbonneau
  \emph{et~al.}}(2015)\citenamefont{Charbonneau, Jin, Parisi, Rainone, Seoane,
  and Zamponi}}]{CJPRSZ15}
\bibinfo{author}{\bibfnamefont{P.}~\bibnamefont{Charbonneau}},
  \bibinfo{author}{\bibfnamefont{Y.}~\bibnamefont{Jin}},
  \bibinfo{author}{\bibfnamefont{G.}~\bibnamefont{Parisi}},
  \bibinfo{author}{\bibfnamefont{C.}~\bibnamefont{Rainone}},
  \bibinfo{author}{\bibfnamefont{B.}~\bibnamefont{Seoane}}, \bibnamefont{and}
  \bibinfo{author}{\bibfnamefont{F.}~\bibnamefont{Zamponi}},
  \bibinfo{journal}{Phys. Rev. E} \textbf{\bibinfo{volume}{92}},
  \bibinfo{pages}{012316} (\bibinfo{year}{2015}).

\bibitem[{\citenamefont{Biroli and P.Urbani}(2017)}]{BU17}
\bibinfo{author}{\bibfnamefont{G.}~\bibnamefont{Biroli}} \bibnamefont{and}
  \bibinfo{author}{\bibnamefont{P.Urbani}} (\bibinfo{year}{2017}),
  \bibinfo{note}{{\tt arXiv:1704.04649}}.

\bibitem[{\citenamefont{Goodrich  \emph{et~al.}}(2012)\citenamefont{Goodrich,
  Liu, and Nagel}}]{GLN12}
\bibinfo{author}{\bibfnamefont{C.~P.} \bibnamefont{Goodrich}},
  \bibinfo{author}{\bibfnamefont{A.~J.} \bibnamefont{Liu}}, \bibnamefont{and}
  \bibinfo{author}{\bibfnamefont{S.~R.} \bibnamefont{Nagel}},
  \bibinfo{journal}{Phys. Rev. Lett.} \textbf{\bibinfo{volume}{109}},
  \bibinfo{pages}{095704} (\bibinfo{year}{2012}).

\bibitem[{\citenamefont{Hicks  \emph{et~al.}}(2017)\citenamefont{Hicks,
  Wheatley, Godfrey, and Moore}}]{Moore:2017}
\bibinfo{author}{\bibfnamefont{C.~L.} \bibnamefont{Hicks}},
  \bibinfo{author}{\bibfnamefont{M.~J.} \bibnamefont{Wheatley}},
  \bibinfo{author}{\bibfnamefont{M.~J.} \bibnamefont{Godfrey}},
  \bibnamefont{and} \bibinfo{author}{\bibfnamefont{M.~A.} \bibnamefont{Moore}}
  (\bibinfo{year}{2017}), \bibinfo{note}{{\tt 1708.05644}}.

\bibitem[{\citenamefont{Seoane  \emph{et~al.}}(2017)\citenamefont{Seoane, Reid,
  de~Pablo, and Zamponi}}]{Bea:2017}
\bibinfo{author}{\bibfnamefont{B.}~\bibnamefont{Seoane}},
  \bibinfo{author}{\bibfnamefont{D.~R.} \bibnamefont{Reid}},
  \bibinfo{author}{\bibfnamefont{J.~J.} \bibnamefont{de~Pablo}},
  \bibnamefont{and} \bibinfo{author}{\bibfnamefont{F.}~\bibnamefont{Zamponi}}
  (\bibinfo{year}{2017}), \bibinfo{note}{{\tt arXiv:1709.04930}}.

\bibitem[{\citenamefont{Andersen  \emph{et~al.}}(1971)\citenamefont{Andersen,
  Weeks, and Chandler}}]{WCA71}
\bibinfo{author}{\bibfnamefont{H.~C.} \bibnamefont{Andersen}},
  \bibinfo{author}{\bibfnamefont{J.~D.} \bibnamefont{Weeks}}, \bibnamefont{and}
  \bibinfo{author}{\bibfnamefont{D.}~\bibnamefont{Chandler}},
  \bibinfo{journal}{Phys. Rev. A} \textbf{\bibinfo{volume}{4}},
  \bibinfo{pages}{1597} (\bibinfo{year}{1971}).

\bibitem[{\citenamefont{Berthier and Tarjus}(2011)}]{BT11}
\bibinfo{author}{\bibfnamefont{L.}~\bibnamefont{Berthier}} \bibnamefont{and}
  \bibinfo{author}{\bibfnamefont{G.}~\bibnamefont{Tarjus}},
  \bibinfo{journal}{J. Chem. Phys.} \textbf{\bibinfo{volume}{134}},
  \bibinfo{pages}{214503} (\bibinfo{year}{2011}).

\bibitem[{\citenamefont{Mari and Kurchan}(2011)}]{MK11}
\bibinfo{author}{\bibfnamefont{R.}~\bibnamefont{Mari}} \bibnamefont{and}
  \bibinfo{author}{\bibfnamefont{J.}~\bibnamefont{Kurchan}},
  \bibinfo{journal}{J. Chem. Phys.} \textbf{\bibinfo{volume}{135}},
  \bibinfo{pages}{124504} (\bibinfo{year}{2011}).

\bibitem[{\citenamefont{Lerner  \emph{et~al.}}(2016)\citenamefont{Lerner,
  D\"uring, and Bouchbinder}}]{LDB16}
\bibinfo{author}{\bibfnamefont{E.}~\bibnamefont{Lerner}},
  \bibinfo{author}{\bibfnamefont{G.}~\bibnamefont{D\"uring}}, \bibnamefont{and}
  \bibinfo{author}{\bibfnamefont{E.}~\bibnamefont{Bouchbinder}},
  \bibinfo{journal}{Phys. Rev. Lett.} \textbf{\bibinfo{volume}{117}},
  \bibinfo{pages}{035501} (\bibinfo{year}{2016}).

\bibitem[{\citenamefont{Mizuno  \emph{et~al.}}(2017)\citenamefont{Mizuno,
  Shiba, and Ikeda}}]{MSI17}
\bibinfo{author}{\bibfnamefont{H.}~\bibnamefont{Mizuno}},
  \bibinfo{author}{\bibfnamefont{H.}~\bibnamefont{Shiba}}, \bibnamefont{and}
  \bibinfo{author}{\bibfnamefont{A.}~\bibnamefont{Ikeda}}
  (\bibinfo{year}{2017}), \bibinfo{note}{{\tt arXiv:1703.10004}}.

\bibitem[{\citenamefont{Keys  \emph{et~al.}}(2011)\citenamefont{Keys, Hedges,
  Garrahan, Glotzer, and Chandler}}]{KHGGC11}
\bibinfo{author}{\bibfnamefont{A.~S.} \bibnamefont{Keys}},
  \bibinfo{author}{\bibfnamefont{L.~O.} \bibnamefont{Hedges}},
  \bibinfo{author}{\bibfnamefont{J.~P.} \bibnamefont{Garrahan}},
  \bibinfo{author}{\bibfnamefont{S.~C.} \bibnamefont{Glotzer}},
  \bibnamefont{and} \bibinfo{author}{\bibfnamefont{D.}~\bibnamefont{Chandler}},
  \bibinfo{journal}{Phys. Rev. X} \textbf{\bibinfo{volume}{1}},
  \bibinfo{pages}{021013} (\bibinfo{year}{2011}).

\bibitem[{\citenamefont{Candelier  \emph{et~al.}}(2010)\citenamefont{Candelier,
  Widmer-Cooper, Kummerfeld, Dauchot, Biroli, Harrowell, and
  Reichman}}]{CWKDBHR10}
\bibinfo{author}{\bibfnamefont{R.}~\bibnamefont{Candelier}},
  \bibinfo{author}{\bibfnamefont{A.}~\bibnamefont{Widmer-Cooper}},
  \bibinfo{author}{\bibfnamefont{J.~K.} \bibnamefont{Kummerfeld}},
  \bibinfo{author}{\bibfnamefont{O.}~\bibnamefont{Dauchot}},
  \bibinfo{author}{\bibfnamefont{G.}~\bibnamefont{Biroli}},
  \bibinfo{author}{\bibfnamefont{P.}~\bibnamefont{Harrowell}},
  \bibnamefont{and} \bibinfo{author}{\bibfnamefont{D.~R.}
  \bibnamefont{Reichman}}, \bibinfo{journal}{Phys. Rev. Lett.}
  \textbf{\bibinfo{volume}{105}}, \bibinfo{pages}{135702}
  (\bibinfo{year}{2010}).

\bibitem[{\citenamefont{Cubuk  \emph{et~al.}}(2015)\citenamefont{Cubuk,
  Schoenholz, Rieser, Malone, Rottler, Durian, Kaxiras, and Liu}}]{CSRMRDKL15}
\bibinfo{author}{\bibfnamefont{E.~D.} \bibnamefont{Cubuk}},
  \bibinfo{author}{\bibfnamefont{S.~S.} \bibnamefont{Schoenholz}},
  \bibinfo{author}{\bibfnamefont{J.~M.} \bibnamefont{Rieser}},
  \bibinfo{author}{\bibfnamefont{B.~D.} \bibnamefont{Malone}},
  \bibinfo{author}{\bibfnamefont{J.}~\bibnamefont{Rottler}},
  \bibinfo{author}{\bibfnamefont{D.~J.} \bibnamefont{Durian}},
  \bibinfo{author}{\bibfnamefont{E.}~\bibnamefont{Kaxiras}}, \bibnamefont{and}
  \bibinfo{author}{\bibfnamefont{A.~J.} \bibnamefont{Liu}},
  \bibinfo{journal}{Phys. Rev. Lett.} \textbf{\bibinfo{volume}{114}},
  \bibinfo{pages}{108001} (\bibinfo{year}{2015}).

\bibitem[{\citenamefont{Jack and Garrahan}(2016)}]{JG16}
\bibinfo{author}{\bibfnamefont{R.~L.} \bibnamefont{Jack}} \bibnamefont{and}
  \bibinfo{author}{\bibfnamefont{J.~P.} \bibnamefont{Garrahan}},
  \bibinfo{journal}{Phys. Rev. Lett.} \textbf{\bibinfo{volume}{116}},
  \bibinfo{pages}{055702} (\bibinfo{year}{2016}).

\bibitem[{\citenamefont{Falk and Langer}(1998)}]{FL98}
\bibinfo{author}{\bibfnamefont{M.}~\bibnamefont{Falk}} \bibnamefont{and}
  \bibinfo{author}{\bibfnamefont{J.}~\bibnamefont{Langer}},
  \bibinfo{journal}{Phys. Rev. E} \textbf{\bibinfo{volume}{57}},
  \bibinfo{pages}{7192} (\bibinfo{year}{1998}).

\bibitem[{\citenamefont{Schall  \emph{et~al.}}(2007)\citenamefont{Schall,
  Weitz, and Spaepen}}]{SWS07}
\bibinfo{author}{\bibfnamefont{P.}~\bibnamefont{Schall}},
  \bibinfo{author}{\bibfnamefont{D.~A.} \bibnamefont{Weitz}}, \bibnamefont{and}
  \bibinfo{author}{\bibfnamefont{F.}~\bibnamefont{Spaepen}},
  \bibinfo{journal}{Science} \textbf{\bibinfo{volume}{318}},
  \bibinfo{pages}{1895} (\bibinfo{year}{2007}).

\bibitem[{\citenamefont{Hentschel  \emph{et~al.}}(2010)\citenamefont{Hentschel,
  Karmakar, Lerner, and Procaccia}}]{HKLP10}
\bibinfo{author}{\bibfnamefont{H.~G.~E.} \bibnamefont{Hentschel}},
  \bibinfo{author}{\bibfnamefont{S.}~\bibnamefont{Karmakar}},
  \bibinfo{author}{\bibfnamefont{E.}~\bibnamefont{Lerner}}, \bibnamefont{and}
  \bibinfo{author}{\bibfnamefont{I.}~\bibnamefont{Procaccia}},
  \bibinfo{journal}{Phys. Rev. Lett.} \textbf{\bibinfo{volume}{104}},
  \bibinfo{pages}{025501} (\bibinfo{year}{2010}).

\bibitem[{\citenamefont{Puosi  \emph{et~al.}}(2016)\citenamefont{Puosi,
  Rottler, and Barrat}}]{PRB16}
\bibinfo{author}{\bibfnamefont{F.}~\bibnamefont{Puosi}},
  \bibinfo{author}{\bibfnamefont{J.}~\bibnamefont{Rottler}}, \bibnamefont{and}
  \bibinfo{author}{\bibfnamefont{J.-L.} \bibnamefont{Barrat}},
  \bibinfo{journal}{Phys. Rev. E} \textbf{\bibinfo{volume}{94}},
  \bibinfo{pages}{032604} (\bibinfo{year}{2016}).

\end{thebibliography}

\end{document}